\documentclass[12pt]{iopart}
\usepackage{iopams}  
\usepackage{cite,hyperref}

\usepackage{graphics,color,xcolor,graphicx}
\usepackage{dsfont} % \(   \mathds{1}   \) identity operator

\begin{document}
	
	\title{Robust optomechanical state transfer under composite phase driving}
	
	\author{C. Ventura-Vel\'{a}zquez}
	\address{Instituto Nacional de Astrof\'{i}sica, \'{O}ptica y Electr\'{o}nica, Calle Luis Enrique Erro No.~1. Sta. Ma. Tonantzintla, Pue. C.P. 72840, M\'{e}xico.}
	
	\author{Benjam\'in Jaramillo \'Avila}
	\address{CONACYT-Instituto Nacional de Astrof\'{i}sica, \'{O}ptica y Electr\'{o}nica, Calle Luis Enrique Erro No.~1. Sta. Ma. Tonantzintla, Pue. C.P. 72840, M\'{e}xico.}
	\ead{jaramillo@inaoep.mx}
	
	\author{Elica Kyoseva}
	\address{Insitute of Solid State Physics, Bulgarian Academy of Sciences, 72~Tsarigradsko Chauss\'{e}e, 1785 Sofia, Bulgaria.}
	\address{School of Physics and Astronomy, Tel Aviv University, Tel Aviv 69978, Israel.}
	
	\author{B. M. Rodr\'iguez-Lara}
	\address{Tecnol\'ogico de Monterrey, Escuela de Ingenier\'ia y Ciencias, Ave. Eugenio Garza Sada~2501, Monterrey, N.L., M\'exico, 64849.}
	\address{Instituto Nacional de Astrof\'{i}sica, \'{O}ptica y Electr\'{o}nica, Calle Luis Enrique Erro No.~1. Sta. Ma. Tonantzintla, Pue. C.P. 72840, M\'{e}xico.}
	
\begin{abstract}
	We propose a technique for robust optomechanical state transfer using phase-tailored composite pulse driving with constant amplitude. Our proposal is inspired by coherent control techniques in lossless driven qubits. 
	We demonstrate that there exist optimal phases for maximally robust excitation exchange in lossy strongly-driven optomechanical state transfer. 
	In addition, our proposed composite phase driving also protects against random variations in the parameters of the system.
	However, this driving can take the system out of its steady state. 
	For this reason, we use the ideal optimal phases to produce smooth sequences that both maintain the system close to its steady state and optimize the robustness of optomechanical state transfer. 
\end{abstract}
	
	\maketitle
	
\section*{Introduction}\label{Sec:Introduction}
The essence of optomechanical systems (OMS) is the coupling between light and mechanical motion. 
Advances in micro and nano fabrication techniques have lead to optical cavities coupled to micro and nano mechanical oscillators, 
where the coupling is provided by the radiation pressure of photons in the optical cavity acting over the mechanical elements. 
These optical cavities are typically pumped by a laser, which serves as a tool to control the system. 
Cavity optomechanical systems evolved from the Fabry-P\'erot cavity to a plethora of devices 
like microtoroids and microresonators \cite{Anetsberger2008}, photonic crystals \cite{Zhang2015}, superconducting microwave circuits \cite{Regal2008,Bernier2018}, ultracold atoms \cite{StamperKurn2014}, among many others \cite{Aspelmeyer2014,Kippenberg2008,Metcalfe2014}.

Optomechanical systems display a range of physical effects that make them a powerful platform for high-precision metrology and quantum-state control. 
They show bistable behavior \cite{Dorsel1983}, 
which is equivalent to that of a Kerr medium \cite{Aldana2013}, 
and display selective transfer over narrow wavelength windows, known as optomechanically-induced transparency \cite{Weis2010,Karuza2013}. 
The latter is equivalent to electromagnetically induced transparency in atoms \cite{Harris1990,Boller1991,Fleischhauer2005} and its plasmonic and metamaterial analogs \cite{Liu2009,Liu2010,Papasimakis2008}. 
In cavity optomechanical systems, 
the motion of the mechanical oscillator can be cooled by tuning the laser that pumps the cavity \cite{Mancini1998,Marquardt2008,Marquardt2007,Liu2013}, 
leading to experiments where a nano-oscillator is cooled to its quantum-mechanical ground state \cite{Chan2011,OConnell2010,Teufel2011}. 
Previous works propose diverse techniques to enhance optomechanical cooling, for example, 
by dynamically modifying the damping\cite{Xunnong2017}, 
using squeezed light\cite{SafaviNaeini2013,Asjad2017,Rakhubovsky2016,Miwa2014}, 
feedback-controlled light\cite{Asjad2016,Rossi2017}, 
or considering the effects of non-Markovian evolution \cite{Triana2016}. 
These developments show that optomechanical effects allow control over quantum optical and mechanical states 
leading to exciting proposals to use these systems as transducers\cite{Stannigel2010,Stannigel2011,SafaviNaeini2011,Stannigel2012,Tian2015,Vostrosablin2017}.

In the following, we review the formalism that describes quantum excitation exchange in strongly-driven optomechanical systems. 
The result is a well-known linearized lossy model. 
Next, we draw from lossless coherent control techniques in qubits \cite{Torosov2011a,Torosov2011b} and extend them to this linearized effective model of optomechanical systems. 
Our proposal relies on constant-amplitude composite phase-dependent pumping to achieve robust optomechanical state transfer. 
This phase-dependent driving produces interference in the evolution of cavity and mechanical quantum states. 
We engineer this interference to minimize the effect of deviations in the parameters that characterize the system, \textit{i.e.} to produce robust optomechanical state transfer. 
For the sake of completeness, we produce a central-limit analysis allowing for random variations in the physical parameters that characterize the optomechanical system and compare results from the standard constant-phase sideband state transfer and our method.
Next, we discuss the effects of composite phase sequences on the semiclassical steady-state of an optomechanical system and show that our original proposal can be used as a recipe to produce more realistic smooth phase sequences with potential for experimental implementation.
We close with our conclusions. 	

\section*{Results}\label{Sec:Results}
\subsection*{Quantum excitation exchange}\label{Sec:Results:Sub:QEE}
First we provide a quick review of quantum excitation exchange. 
Our starting point is the standard Hamiltonian describing a laser-driven optomechanical system in a frame rotating at the pump laser frequency \cite{Pace1993,Law1995}, 
\begin{eqnarray}
\label{Eq:H0}
\hat{H}_{0}
&=&
\left(\omega_{c}-\omega_{p}\right) \hat{a}^{\dag}\hat{a}
+\omega_{m}\, \hat{b}^{\dag}\hat{b}
+g_{0}\, \hat{a}^{\dag}\hat{a}
\left(
\hat{b}^{\dag}+\hat{b}
\right)
+i\,\frac{\varepsilon}{2}
\left(
e^{i\,\phi(t)}\,\hat{a}^{\dag}
-
e^{-i\,\phi(t)}\,\hat{a}
\right),
\end{eqnarray}
where the cavity (mechanical) field mode frequency is $\omega_{c}$ ($\omega_{m}$) and its annihilation operator is $\hat{a}$ ($\hat{b}$), the optomechanical coupling is $g_{0}$, the driving-laser strength is $\varepsilon$, and its frequency is $\omega_{p}$. 
We focus on the effect of phase sequences: The system is driven by a constant-power laser where its phase, $\phi(t)$, is given by a piecewise function which takes constant values in each of its pieces. 
For the sake of space, we use the shorthand $\phi \equiv \phi(t)$.

It is well known that strong driving allows us to split the dynamics into semi-classical and quantum fluctuation components, $\hat{a}= \alpha e^{i\phi} + \hat{c} $ and $\hat{b}= \beta + \hat{d}$ \cite{Mancini1994,Paternostro2006}. The semi-classical steady state for the cavity (mechanical) mode has a coherent amplitude $\alpha= \varepsilon /\left[\kappa + i\,2\left(\omega_{c}-\omega_{p}\right)\right]$ ( $\beta= -i\,g_{0}|\alpha|^{2} /\left[\gamma/2 +i\,\omega_{m}\right]$), where the cavity (mechanical) decay rate is given by $\kappa$ ($\gamma$). The semi-classical boson numbers for cavity and mechanical modes are respectively $n_{p} = \vert \alpha \vert^{2}$ and $n_{m}=\vert \beta \vert^{2}$. This mean field approximation allows the linearization of the optomechanical interaction and leads to a Hamiltonian for the quantum components,
\begin{eqnarray}
\label{Eq:LinearizedHamiltonian}
\hat{H}
&=&
-\Delta\, \hat{c}^{\dag}\hat{c}
+
\omega_{m}\, \hat{d}^{\dag}\hat{d}
+
g\left(
e^{i\,\varphi}\, \hat{c}^{\dag}
+
e^{-i\,\varphi}\, \hat{c}
\right)
\left(
\hat{d}^{\dag} + \hat{d}
\right),
\end{eqnarray} 
where we define a detuning $\Delta= \omega_{p} -\omega_{c} - 2\,g_{0}\, \textrm{Re}\left(\beta\right)$ and an enhanced optomechanical coupling $g= g_{0} |\alpha|$ \cite{Genes2008}. The auxiliary phase, $\varphi \equiv \varphi(t) = \phi + \arg\left(\alpha\right)$ inherits the time dependence from the driving laser phase.

Under red-detuned driving, $\Delta= -\omega_{m}$, a rotating wave approximation leads to beam-splitter-like interaction terms, $e^{i\varphi}\,\hat{c}^{\dag}\, \hat{d} + e^{-i\varphi}\,\hat{c}\, \hat{d}^{\dag}$. 
This interaction allows for quantum excitation exchange between cavity and mechanical modes. 
The full dynamics is described by the corresponding quantum Langevin equations (QLE) \cite{Gardiner1985,Walls2008},
\begin{eqnarray}
%	\label{Eq:LangevinFull}
\partial_{t}
\left(\begin{array}{c}
\hat{c}\\
\hat{d}
\end{array}\right)
=
\left(\begin{array}{cc}
-\left(i\,\omega_{m}+\kappa/2\right)	&	-i\,g\,e^{i\,\varphi}
\\
-i\,g\,e^{-i\,\varphi}	&	-\left(i\,\omega_{m}+\gamma/2\right)
\end{array}\right)
\left(\begin{array}{c}
\hat{c}\\
\hat{d}
\end{array}\right)
-\left(\begin{array}{c}
\sqrt{\kappa}\,\hat{\xi}_{c}\\
\sqrt{\gamma}\,\hat{\xi}_{m}
\end{array}\right),
\end{eqnarray}
where the quantum Gaussian noise for the cavity (mechanical) mode is described by the operator $\hat{\xi}_{c}$ ($\hat{\xi}_{m}$) with correlation functions $\left\langle \hat{\xi}_{\left(c,m\right)}^{\dag}\left(t\right) \hat{\xi}_{\left(c,m\right)}\left(s\right)
\right\rangle =	n_{th}^{\left(c,m\right)}\,	\delta\left(t-s\right)$ and $\left\langle \hat{\xi}_{\left(c,m\right)}\left(t\right) \hat{\xi}_{\left(c,m\right)}^{\dag}\left(s\right)\right\rangle= \left( n_{th}^{\left(c,m\right)}+1\right)\,\delta\left(t-s\right)$. The parameters $n_{th}^{\left(c,m\right)}$ are the average occupation numbers in thermal equilibrium for the baths. 
We define the constants
$\Gamma= \left(\kappa-\gamma\right) / \left(4\, g\right)$ and 
$\mu= \left(\kappa+\gamma\right) /4$, and introduce the vectors 
$\vec{r}= e^{\left( \mu+i\,\omega_{m} \right)t}\, \left(\hat{c},\, \hat{d}\right)^{T}$ and 
$\vec{r}_{\mathrm{in}}= e^{\left( \mu+i\,\omega_{m} \right)t}\, \left(\sqrt{\kappa}\,\hat{\xi}_{c},\, \sqrt{\gamma}\,\hat{\xi}_{m}\right)^{T}$ to rewrite the QLE,
\begin{eqnarray}
%	\label{Eq:LangevinBrief}
-\partial_{t} \vec{r}
=
i\, g\, \hat{\mathcal{H}}\, \vec{r}
+
\vec{r}_{\mathrm{in}},
\end{eqnarray}
where $\hat{\mathcal{H}}$ is a $2\times2$ non-Hermitian matrix,
\begin{eqnarray}
%	\label{Eq:HamiltonianLangevin}
\hat{\mathcal{H}}
=
\left(
e^{i\, \varphi}\, \hat{\sigma}_{+}
+
e^{-i\, \varphi}\, \hat{\sigma}_{-}
\right)
-
i\, \Gamma\, \hat{\sigma}_{z},
\end{eqnarray}
and we used the standard definition for Pauli matrices. 	
When the driving-laser phase is constant, it is straightforward to calculate the evolution of $\vec{r}(\tau)$ for the interval $\tau = t-t_{0}$,
\begin{eqnarray}
\label{Eq:LangevinSolution}
\vec{r}\left( \tau \right)
=\,
\hat{U}_{\varphi}\left(\tau \right)\,
\vec{r}\left(t_{0}\right)
-\int_{t_{0}}^{t}
\hat{U}_{\varphi}\left(t-y\right)\,
\vec{r}_{\mathrm{in}}\left(y\right)\,
dy,
\end{eqnarray}
where $\hat{U}_{\varphi}$ is the constant-phase evolution matrix due to the Hamiltonian $\hat{\mathcal{H}}$,
\begin{eqnarray}
\label{Eq:UVarphi}
\hat{U}_{\varphi}\left(\tau\right)
=
e^{-i\,g\, \hat{\mathcal{H}}\, \tau }
=
\cos\left(\Omega\,g\,\tau \right)\,
\mathds{1}_{2}
-
\frac{i}{\Omega}\,\sin\left(\Omega\,g\,\tau \right)\,
\hat{\mathcal{H}}.
\end{eqnarray}
Here, we use the notation $\mathds{1}_{2}$ for the $2\times2$ identity matrix and introduce an effective Rabi frequency $\Omega= \sqrt{1-\Gamma^{2}}$ that is real for small losses compared to the coupling parameter $(\kappa - \gamma ) \ll g$. 
The constant-phase evolution, Eq.~(\ref{Eq:UVarphi}), suggests a time scale $\tau_{0}= \pi/\left(2\,g\,\Omega\right)$, which corresponds to an accumulated interaction area $A(\tau_{0})= \pi/2$ for quantum excitation exchange. 	
The mean photon and phonon numbers, with constant-phase driving,
\begin{eqnarray}
\left\langle 
\hat{c}^{\dag}\hat{c}
\right\rangle \left(\tau\right)
&=&
\left(
n_{0}\,
|U_{11}\left(\tau\right)|^{2}
+
m_{0}\,
|U_{12}\left(\tau\right)|^{2}
\right) e^{-2\,\mu\,t}
+
\tilde{\kappa}\, S_{11}\left(\tau\right)
+
\tilde{\gamma}\, S_{12}\left(\tau\right),
\nonumber \\
\left\langle 
\hat{d}^{\dag}\hat{d}
\right\rangle \left(\tau\right)
&=&
\left(
n_{0}\,
|U_{21}\left(\tau\right)|^{2}
+
m_{0}\,
|U_{22}\left(\tau\right)|^{2}
\right) e^{-2\,\mu\,t}
+
\tilde{\kappa}\, S_{21}\left(\tau\right)
+
\tilde{\gamma}\, S_{22}\left(\tau\right), 	\label{Eq:NumbersExpectation}
\end{eqnarray}
depend on the initial-time mean photon and phonon occupation numbers, $n_{0}$ and $m_{0}$ respectively, 
and on the decay rates modified by their respective thermal boson numbers, $\tilde{\kappa}= \kappa\, n_{th}^{\left(c\right)}$ and $\tilde{\gamma}= \gamma\, n_{th}^{\left(m\right)}$. 
The functions $U_{jk}\left(\tau\right) \equiv \left[ U_{\varphi}(\tau) \right]_{j,k}$ are a shorthand notation for the matrix elements of the evolution operator,	and the functions $S_{jk}\left(\tau\right) \equiv \int_{0}^{\tau} e^{-2\,\mu\,x} |U_{jk}\left(x\right)|^{2} dx$ arise from quantum noise correlations.

\begin{figure}[tpb!]
	\centering
	\includegraphics{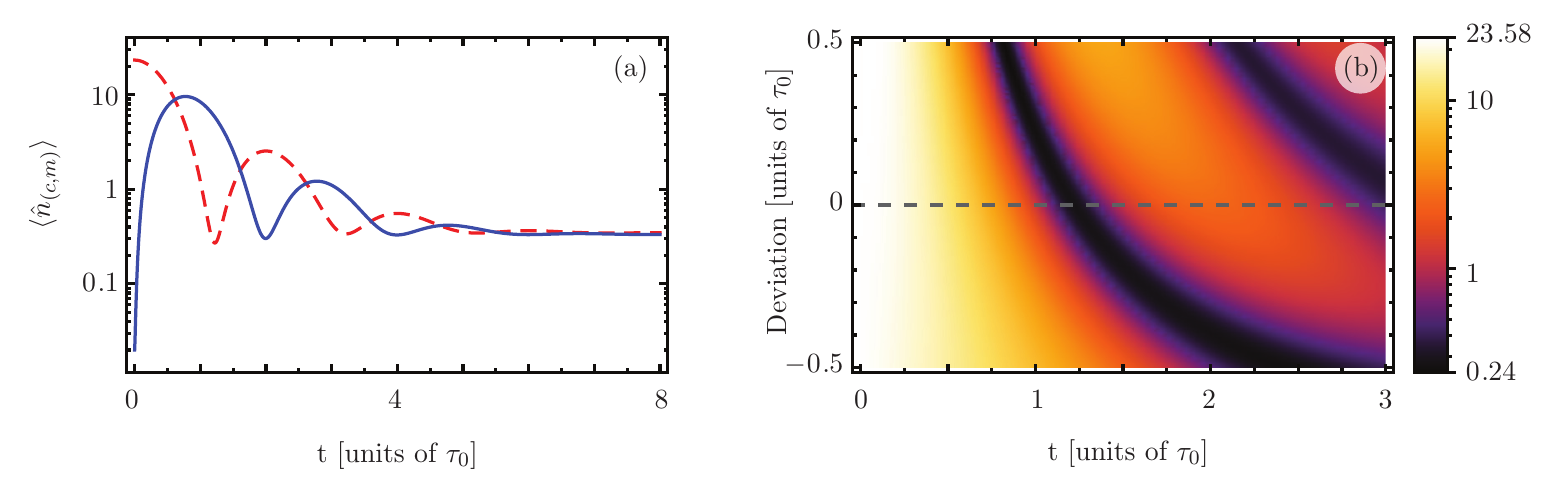}
	\caption{Time evolution of (a) the expectation values for the quantum photon (solid blue) and phonon (dashed red) numbers and (b) quantum expectation value for the phonon number considering deviations from the ideal accumulated interaction area $A(\tau_{0})= \pi/2$ (gray dashed line).}\label{Fig:Oscillations}
\end{figure}

Figure \ref{Fig:Oscillations}(a) shows the well-known oscillatory quantum excitation exchange between cavity and mechanical modes under continuous-wave driving and losses. Figure \ref{Fig:Oscillations}(b) shows the effect of measuring the mean phonon number when the accumulated interaction area deviates from the $\pi/2$ value required to obtain the excitation exchange; the dashed line shows the ideal case. 
The parameters in these calculations are 
$g_{0}/\omega_{m}= 1.887\times 10^{-5}$, 
$\kappa/\omega_{m}\approx 1.119\times 10^{-2}$, 
$\gamma/\omega_{m}= 9.434\times 10^{-6}$, 
$n_{p}= 180\times 10^{3}$, 
$n_{0}=0.02$, 
and $m_{0}=23.25$. 
The temperature of the bath is $T= 25 \, mK$, equivalent to $n_{th}^{\left(m\right)} \approx 32.27$ thermal phonons and a negligible number of thermal photons. However, the strong driving of the cavity and the experimental setup allows the cavity to reach thermal equilibrium at a higher occupancy number $n_{th}^{\left(c\right)}\approx 0.305$. These experimental values and considerations were retrieved from Lecocq \textit{et al.}\cite{Lecocq2015}. We introduce an area deviation, $A = A(\tau_{0}) + \delta A $, by changing the ideal enhanced coupling, $g$. 
When the effective interaction area is larger than the target, $A>\pi/2$, the oscillation period shortens, and when it is smaller, $A<\pi/2$, the period increases, see Fig. \ref{Fig:Oscillations}(b). 
In the following, we will show robust excitation exchange, against deviations from the ideal accumulated interaction area, using driving that is constant in amplitude but has a piecewise phase structure.

\subsection*{Robust transfer via composite phase sequences}\label{Sec:Results:Sub:PhaseSequence}
We note that the quantum Langevin equations for the cavity and mechanical excitations, $\hat{c}$ and $\hat{d}$, are similar to those describing a qubit driven by an external coherent electromagnetic field. 
Then, the well-known Rabi oscillations are observed where the state of the qubit is controllably switched between 0 and 1 when the driving is resonant. However, quantum state preparation based on Rabi oscillations is imprecise in the presence of the so-called systematic errors in the system such as inaccuracies in the field magnitude, frequency, duration, coupling strength, and others. 
Composite pulses are a very useful tool to overcome the sensitivity to errors and to realize robust qubit state preparation. A composite pulse is a train of identical pulses (same frequency, strength, and duration) where each has a different, fixed phase. Such composite pulse sequences have been shown to realize a robust qubit rotation for a closed lossless system \cite{Torosov2011a,Torosov2011b,Kyoseva2014}.
In this section, we design a composite pulse sequence to induce robust phonon-photon excitation exchange for lossy optomechanical state transfer given that there are less photons than phonons in the initial state, $n_0 < m_0$. 

We assume a phase-tailored composite pulse sequence with constant driving amplitude. Each constituent pulse in the sequence has a constant phase and a time duration of $\tau_0$ chosen such that the accumulated interaction area is $\pi/2$. We also require that the first and last phases of the sequence are equal to each other. In consequence, the shortest possible sequence has three interactions and a total accumulated interaction area of $3\,\pi/2$. 
The phases of the first and last interactions of the sequence are fixed, but the phase of the middle interaction is a free parameter that we optimize to stabilize the state transfer of mechanical oscillations with respect to deviations from ideal accumulated interaction areas. 
Under these conditions, the final state of the system, at $3\,\tau_0$, is provided by the vector,
\begin{eqnarray}
\vec{r}\left(3\,\tau_{0}\right)
&=&
\hat{U}^{\,\left(3\right)}(\tau_{0}) \,
\vec{r}_{0}
-
\int_{0}^{\tau_{0}}
\hat{U}^{\,\left(3\right)}\left(\tau_0\right) \,
\hat{U}_{0}\left(-z\right) \,
\vec{r}_{\mathrm{in}}\left(z\right) \,
dz
\nonumber
\\
& &
-
\int_{\tau_{0}}^{2\,\tau_{0}}
\hat{U}_{0}\left(\tau_{0}\right) \,
\hat{U}_{\varphi}\left(2\tau_{0}-z\right) \,
\vec{r}_{\mathrm{in}}\left(z\right)\,
dz
-
\int_{2\,\tau_{0}}^{3\,\tau_{0}}
\hat{U}_{\varphi}\left(3\tau_{0}-z\right) \,
\vec{r}_{\mathrm{in}}\left(z\right) \,
dz,
\label{Eq:r3}
\end{eqnarray}
where $\hat{U}^{\,\left(3\right)}\left(\tau_{0}\right)$
is a transfer matrix for the three-interaction composite sequence, 
\begin{eqnarray}
\label{Eq:U3}
\hat{U}^{\,\left(3\right)}\left(\tau_{0}\right)= \hat{U}_{0}\left(\tau_{0}\right)\, \hat{U}_{\varphi}\left(\tau_{0}\right)\, \hat{U}_{0}\left(\tau_{0}\right),
\end{eqnarray}
where the three $\hat{U}$ matrices describe the evolution of the optomechanical system under different driving phases.
This matrix product produces interference between driving phases, and we exploit this interference to produce robust optomechanical state transfer.

We calculate the mean phonon number, after application of the phase-tailored composite sequence, by using the vector $\vec{r}\left(3\tau_0\right)$ in Eq.~(\ref{Eq:r3}). 
The mean phonon number can be separated in two contributions, 
one with lossy oscillations and one with thermal noise contributions, 
\begin{eqnarray}
\label{Eq:MeanPhononWithPhase}
\left\langle 
\hat{d}^{\dag}\hat{d}
\right\rangle 
\left(3\tau_{0}\right)
=
n_{\mathrm{osc}}^{\left(m\right)}
\left(3\tau_{0}\right)
+
n_{\mathrm{noise}}^{\left(m\right)}
\left(3\tau_{0}\right).
\end{eqnarray}
The oscillatory part depends on the initial mean photon and phonon numbers, $n_0$ and $m_0$ respectively, 
and is independent of the thermal occupation numbers, $n_{th}^{\left(c\right)}$ and $n_{th}^{\left(m\right)}$. 
In contrast, the thermal noise part is independent of $n_0$ and $m_0$ but depends on the thermal occupation numbers through the modified decay rates 
$\tilde{\kappa}= \kappa\, n_{th}^{\left(c\right)}$ and $\tilde{\gamma}= \gamma\, n_{th}^{\left(m\right)}$. 
The oscillatory part is responsible for the excitation exchange, which is similar to the qubit Rabi oscillations. 
It has the form 
\begin{eqnarray}
n_{\mathrm{osc}}^{\left(m\right)}
\left(3\tau_{0}\right) =			e^{-6\,\mu\,\tau_{0}}
\left[
n_{0}\,
|U_{21}^{\,\left(3\right)}\left(\tau_{0}\right)|^{2}
+
m_{0}\,
|U_{22}^{\,\left(3\right)}\left(\tau_{0}\right)|^{2}
\right],
\end{eqnarray}
where the matrix elements of $\hat{U}^{\,\left(3\right)}\left(\tau_{0}\right)$ can be written in terms of the matrix $U_{\varphi}\left(\tau_0\right)$, in Eq.~(\ref{Eq:UVarphi}), with $\varphi=0$,
\begin{eqnarray}
U_{11}^{\,\left(3\right)} \left(\tau_{0}\right)
&=&
\left[ U_{11}(\tau_{0}) \right]^{3}
+
U_{12}(\tau_{0})\,U_{21}(\tau_{0})
\left[
U_{22}(\tau_{0})
+
2\,U_{11}(\tau_{0})\,\cos\,\varphi
\right],
\nonumber \\
U_{12}^{\,\left(3\right)} \left(\tau_{0}\right)
&=&
U_{12}(\tau_{0})
\left\{
\left[ U_{11}(\tau_{0}) \right]^{2} + \left[ U_{22}(\tau_{0}) \right]^{2}
+
e^{-i\,\varphi}\, U_{12}(\tau_{0})\,U_{21}(\tau_{0})
+
e^{i\,\varphi}\, U_{11}(\tau_{0})\,U_{22}(\tau_{0})
\right\},
\nonumber \\
U_{21}^{\,\left(3\right)} \left(\tau_{0}\right)
&=&
U_{21}(\tau_{0})
\left\{
\left[ U_{11}(\tau_{0}) \right]^{2} + \left[ U_{22}(\tau_{0}) \right]^{2}
+
e^{i\,\varphi}\, U_{12}(\tau_{0})\,U_{21}(\tau_{0})
+
e^{-i\,\varphi}\, U_{11}(\tau_{0})\,U_{22}(\tau_{0})
\right\},
\nonumber \\
U_{22}^{\,\left(3\right)} \left(\tau_{0}\right)
&=&
\left[ U_{22}(\tau_{0}) \right]^{3}
+
U_{12}(\tau_{0})\,U_{21}(\tau_{0})
\left[
U_{11}(\tau_{0})
+
2\,U_{22}(\tau_{0})\,\cos\,\varphi
\right].
\end{eqnarray}
The quantum noise part has a more complicated expression, which we omit in the interest of space. 
In order to achieve optimal excitation exchange, we need to minimize the component $|U_{22}^{\,\left(3\right)}\left(\tau_{0}\right)|^{2}$ around the single-interaction accumulated area $\pi/2$, or, alternatively, maximize $|U_{21}^{\,\left(3\right)}\left(\tau_{0}\right)|^{2}$. 
By minimizing the derivative of 
$|U_{22}^{\,\left(3\right)}\left(\tau_{0}\right)|^{2}$ 
with respect to the accumulated interaction area, we obtain a condition for the optimal phase, 
\begin{eqnarray}
\cos\left(\varphi_\mathrm{opt}\right)
=
\frac{3\,\Gamma^{2}-1}{2}, \label{eq:optangle}
\end{eqnarray} 
which for a lossless system yields $\varphi_{\mathrm{opt}}= \pm 2\pi/3$. 
The latter is consistent with the result reported by Torosov \textit{et al.}\cite{Torosov2011a,Torosov2011b} 
for a qubit coherently driven on resonance and without losses.

\begin{figure}[tpb!]
	\centering
	\includegraphics{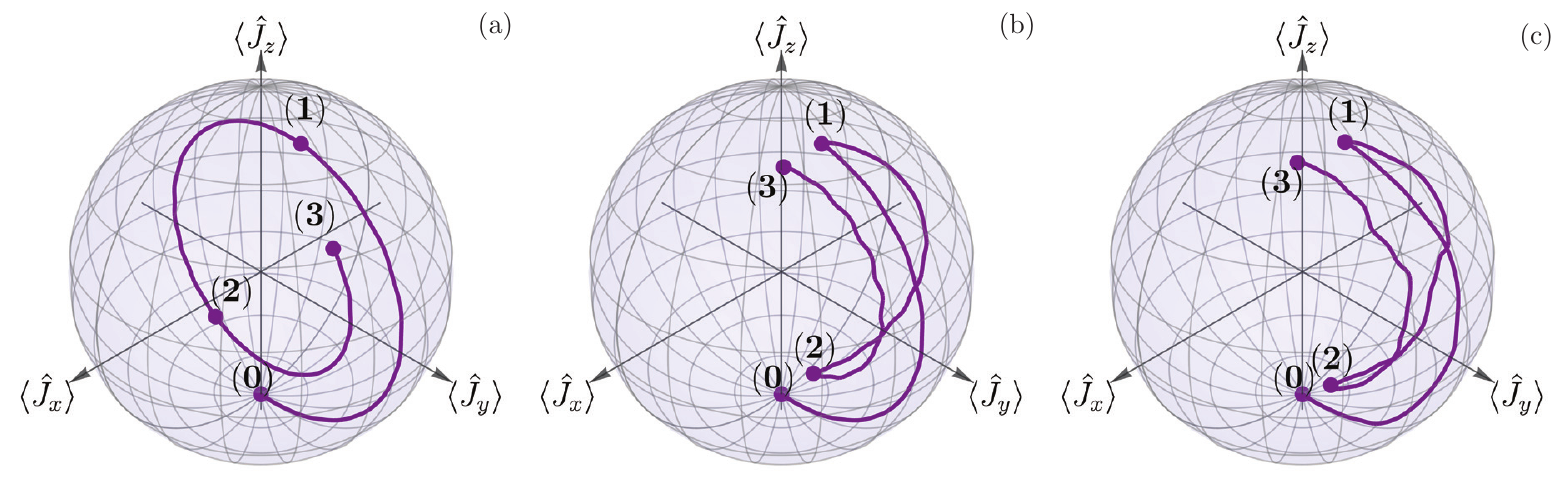}
	\caption{Angular momentum representation of optomechanical state transfer. 
		(a) Displays a three-interaction evolution with a $-10\%$ deviation and constant driving phase. 
		(b) Displays the same evolution under a composite driving sequence that maximizes robustness. 
		(c) Displays the same evolution as (b), but where additional time-dependent white-noise fluctuations are added on 
		$g$, $\omega_{m}$, $\Delta$, $\kappa$, and $\gamma$.}\label{Fig:Spheres}
\end{figure}

In order to introduce visual cues, physical insight, and show that the derived optimal phase sequence produces robust transfer, 
we numerically calculate the evolution using the linearized optomechanical model, Eq.~(\ref{Eq:LinearizedHamiltonian}), 
under Lindblad's phenomenological approach in Fig.~\ref{Fig:Spheres}.
We use Schwinger's two-oscillator representation of angular momentum,
$\hat{J}_{x} = \left( \hat{c}^{\dagger} \hat{d} + \hat{c} \hat{d}^{\dagger} \right) / 2$,
$\hat{J}_{y} =-i\left( \hat{c}^{\dagger} \hat{d} - \hat{c} \hat{d}^{\dagger} \right) / 2$, and
$\hat{J}_{z} = \left( \hat{c}^{\dagger} \hat{c} - \hat{d}^{\dagger} \hat{d} \right)/ 2 $, 
to visualize the effect of a three-interaction composite sequence with the optimal phase derived from Langevin approach, Eq.~(\ref{eq:optangle}).
In all cases shown, the initial state has a single quanta in the mechanical oscillator and vacuum in the cavity. 
The baseline values of the parameters are 
$g/\omega_{m} = 5\times10^{-2}$, 
$\kappa/\omega_{m} = 4\times10^{-3}$, 
$\gamma/\omega_{m} = 8\times10^{-3}$,
$n_{th}^{c}=n_{th}^{m}=0$, 
and $\Delta = -\omega_{m}$. 
Here, we introduce a deviation of $-10\%$; that is, the switching time is $0.9 ~\tau_{0}$ instead of the ideal transfer time $\tau_0$. 
Figure \ref{Fig:Spheres}(a) shows evolution under constant driving. 
We can see that the effect of measuring before complete state transfer in the evolution from the initial state (0) to the end of the first incomplete interval (1).
Figure \ref{Fig:Spheres}(b) shows the same evolution but where the driving phase is changed by $\varphi_\mathrm{opt}$ during the second interaction. 
We can see that the first leg of the interaction is identical to that in Fig.~\ref{Fig:Spheres}(a). 
Then, the phase induces a change of meridian in the trajectory that takes us to a different point (2) in the sphere.
Returning to the original phase, the system performs its third interaction and arrives 
to a state that is closer, to a pure cavity oscillation, that what we would expect from the constant-phase state transfer in Fig. \ref{Fig:Spheres}(a). 
Figure \ref{Fig:Spheres}(c) adds time-dependent random white noise to all parameters in the simulation, 
$g$, $\omega_{m}$, $\Delta$, $\kappa$, and $\gamma$.
This is done by generating a sequence of 50 random numbers in the range from 0.95 to 1.05 for each of the parameters. 
This sequence is then used to generate a smooth interpolated function that multiplies the baseline value of each parameter. 
As we can see, even in the presence of a wrong measuring time, time-dependent random noise in the parameters of the system, and evolution under Lindblad phenomenological master equation, we obtain robust state transfer after the composite phase sequence is applied, Fig.~\ref{Fig:Spheres}(c). 

\begin{figure}[tpb!]
	\centering
	\includegraphics{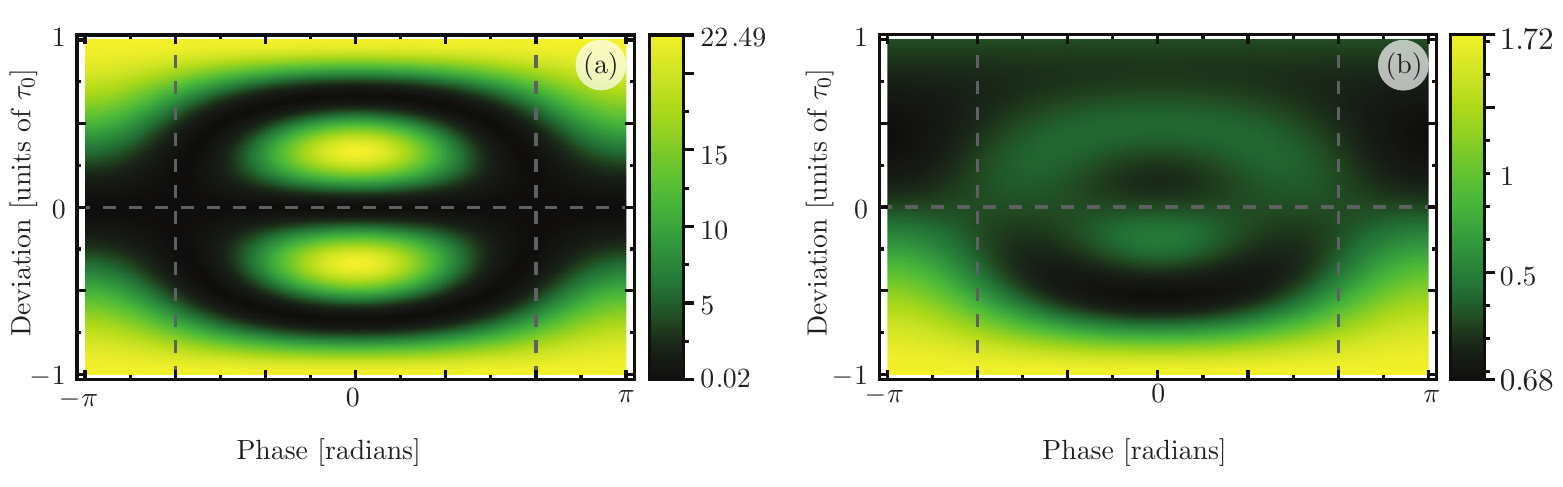}
	\caption{(a) Coherent and (b) thermal noise contributions to the phonon number as a function of the deviation from ideal interaction width and driving phase at time $t_{f} = 3 \tau_{0}$ for parameter values: $g_{0}/\omega_{m}=1.54\times 10^{-4}$, $\kappa/\omega_{m}=1.15\times 10^{-4}$, $\gamma/\omega_{m}=5.36\times 10^{-4}$, $n_{p}=90\times 10^{3}$, $\left(n_{0},\, m_{0}\right)= \left(0.02,\, 23.25\right)$ and $n_{th}^{\left(c,m\right)}=\left(0.21,\,32\right)$ \cite{Cohen2015}.}\label{Fig:MechPhase}
\end{figure}

\begin{figure}[tpb!]
	\centering
	\includegraphics{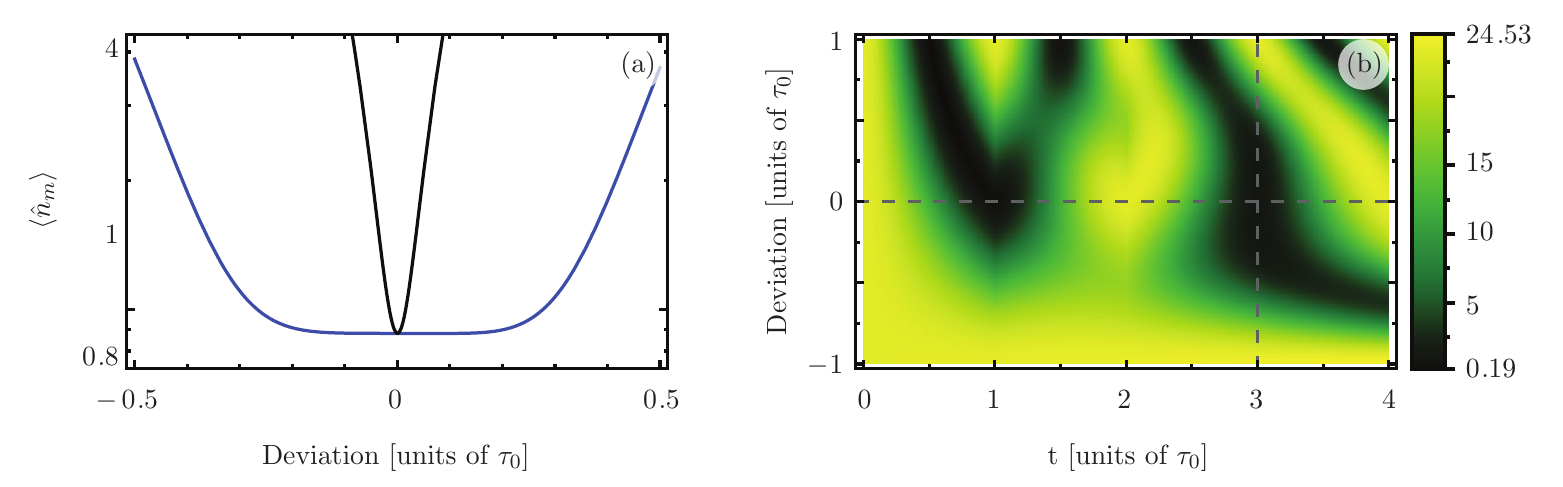}
	\caption{(a) Phonon number expectation value after the composite sequence for driving phases $\varphi = 0$ (black), $\varphi = \pm2\pi/3$ (blue). (b) Time evolution of the phonon number expectation value as a function of the deviation from ideal parameters with a driving phase $\varphi=2\pi/3$. The optimal phases $\varphi_{\mathrm{opt}} = \pm 2\pi/3$ are signaled by vertical dashed lines. Parameter values are equal to those in Fig.~\ref{Fig:MechPhase}.}\label{Fig:Evol}
\end{figure}

Now, let us come back to the analysis of the oscillatory and thermal noise contributions to the mean phonon number in Eq.~(\ref{Eq:MeanPhononWithPhase}). 
In Figure \ref{Fig:MechPhase}, we consider deviations in the accumulated interaction area and the whole range of values for the phase parameter $\varphi$. 
For these physical parameters, the lossy optimal phase is almost identical to the lossless one, $\varphi_{\mathrm{opt}} \approx \pm 2\pi/3$ within $10^{-3}\,\%$. 
The vertical dashed lines show the phases that produce optimally robust state transfer. 
For photon and phonon numbers, the oscillatory, Fig. \ref{Fig:MechPhase}(a), and thermal noise, Fig. \ref{Fig:MechPhase}(b), contributions are invariant to sign exchange in the phase parameter. 
The effect of these behaviors is seen in Fig. \ref{Fig:Evol}(a), where we add oscillatory and thermal noise contributions to get the total phonon number. Figure \ref{Fig:Evol}(b) shows the full time evolution for the phonon number under the optimal composite sequence considering deviations in the accumulated interaction area. 
There is a minimum phonon number region at the intersection between the ideal parameter set (horizontal dashed line) and the endtime of the sequence (vertical dashed line).

\begin{figure}[tpb!]
	\centering
	\includegraphics{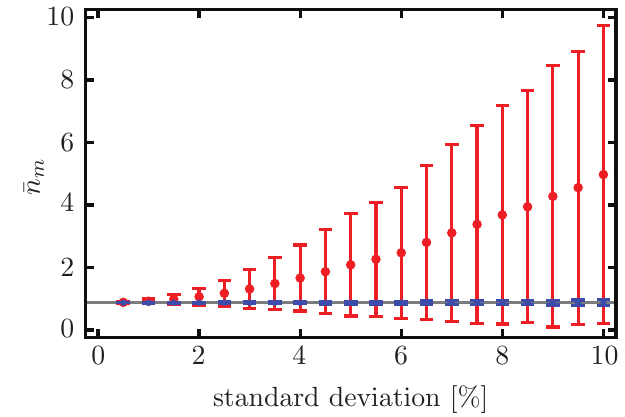}
	\caption{
		Mean value (dots) and standard deviation (bars) of samples of phonon number expectation value samples after three interactions, $\bar{n}_{m}$. Again, each of the physical parameters, $g,\kappa,\gamma$, is generated using a normal distribution with standard deviation determined as a percentage of the central value of the corresponding parameter. 
		Red points and bars correspond to constant-phase evolution and 
		blue points and bars correspond to evolution under composite phase driving and 
		Here, the experimental values are the same as those in Fig.~\ref{Fig:Oscillations} and Lecocq \textit{et al.}~\cite{Lecocq2015}. 
		We used a sample consisting of 5000 instances for each point.
	}\label{Fig:GaussDevs}
\end{figure}

In addition, we perform a central-limit statistical analysis to validate whether or not our composite phase driving proposal
produces robust optomechanical state transfer under random variations in its physical parameters.
In the following, the mean values and standard deviations are calculated from the resulting quantum mean phonon value samples and do not correspond to the quantum average mean phonon number and its uncertainty. 
For our analysis, we calculate a large sample of simultaneous random variations on the physical parameters $g, \kappa, \gamma$, each of them under a normal distribution with fixed mean and standard deviation. 
Next, we calculate the phonon-number expectation value after three interactions, $\langle \hat{n}_m \left(3\tau_0\right)\rangle$, for each of these sets. 
We use two scenarios: 
Evolution under the standard red-sideband state transfer and our composite phase driving proposal. 
While the parameters $g, \kappa, \gamma$ have random variations, the driving sequence is determined by the central values that characterize the system and, therefore, it is constant and does not suffer from random variations. 	
Table~\ref{tab:deviations} collects the mean value of the phonon number and its standard deviation for samples calculated with experimental parameters reported in the literature. 
We find that composite driving produces samples that are more robust against Gaussian deviations in their physical parameters; 
\textit{i.e.} the standard deviation of the sample of phonon numbers is much smaller under our custom phase driving than under constant-phase driving. 
In addition, the mean phonon number is centered around a minimum, 
which means that random variations are more likely to produce an increase in the average phonon number than a decrease.
Table~\ref{tab:deviations} and Figure~\ref{Fig:GaussDevs} show that, as the standard deviation for the random variation of the physical parameters $g, \kappa, \gamma$ increases, the average phonon numbers increase, 
but the increase is more dramatic under constat-phase driving.

\begin{table}
	\begin{center}
		
		\begin{tabular}{||c||c|c||c|c||c|c||}
			\hline 
			&\multicolumn{2}{c||}{\textbf{Cohen \textit{et al.}}\cite{Cohen2015}}
			&\multicolumn{2}{c||}{\textbf{Lecocq \textit{et al.}}\cite{Lecocq2015}}
			&\multicolumn{2}{c||}{\textbf{Gr\"oblacher \textit{et al.}}\cite{Groblacher2009}} 
			\\ \hline
			& Constant		& Composite		& Constant		& Composite		& Constant		& Composite			
			\\ 	\hline
			\multicolumn{7}{c}{\textbf{central values (no random variation)}}
			\\ 	\hline
			$\langle \hat{n}_m \rangle$
			& 0.871			& 0.870			& 0.337			& 0.627			& 0.0454		& 0.0456
			\\ 	\hline
			\multicolumn{7}{c}{\textbf{1\% $\sigma$ in random variations in $g$, $\kappa$, $\gamma$}}
			\\ 	\hline
			$\bar{n}_m$
			& 0.917			& 0.870			& 0.340			& 0.627			& 0.0950		& 0.0456
			\\
			$\sigma$
			& 0.0665		& 0.00845		& 0.0305		& 0.0128		& 0.0705		& 0.000323
			\\ 	\hline
			\multicolumn{7}{c}{\textbf{2\% $\sigma$ in random variations in $g$, $\kappa$, $\gamma$}}
			\\ 	\hline
			$\bar{n}_m$
			& 1.06			& 0.870			& 0.346			& 0.629			& 0.247			& 0.0456
			\\
			$\sigma$
			& 0.274			& 0.0173		& 0.0614		& 0.0263		& 0.289			& 0.000653
			\\ 	\hline
			\multicolumn{7}{c}{\textbf{5\% $\sigma$ in random variations in $g$, $\kappa$, $\gamma$}}
			\\ 	\hline
			$\bar{n}_m$
			& 2.06			& 0.871			& 0.385			& 0.638			& 1.20			& 0.0456
			\\
			$\sigma$
			& 1.592			& 0.0420		& 0.1474		& 0.0718		& 1.481			& 0.00164\\
			\hline
		\end{tabular}
		\caption{
			Mean value, $\bar{n}_m$, and standard deviation, $\sigma$, of phonon number expectation value samples resulting from randomly-variated parameters $g$, $\kappa$, and $\gamma$. 
			We use three sets of central values, 
			Cohen \textit{et al.}\cite{Cohen2015}, 
			with $g/\omega_{m}=4.62\times 10^{-2}$, $\kappa/\omega_{m}=1.15\times 10^{-4}$, $\gamma/\omega_{m}=5.36\times 10^{-4}$;	
			Lecocq \textit{et al.}\cite{Lecocq2015}, 
			with $g/\omega_{m}=8.01\times 10^{-3}$, $\kappa/\omega_{m}=1.12\times 10^{-2}$, $\gamma/\omega_{m}=9.43\times 10^{-6}$;
			and Gr\"oblacher \textit{et al.}\cite{Groblacher2009}, 
			with $g/\omega_{m}=0.434$, $\kappa/\omega_{m}=2.27\times 10^{-3}$, $\gamma/\omega_{m}=1.48\times 10^{-4}$;
			We set the following initial values $n_0=0.01$, $m_0=23.25$, $n_{th}^{(c)}=0.21$, $n_{th}^{(m)}=32.0$ and use 3000 instances in each sample.
		}\label{tab:deviations}
	\end{center}
\end{table}

\begin{figure}[tpb!]
	\centering
	\includegraphics{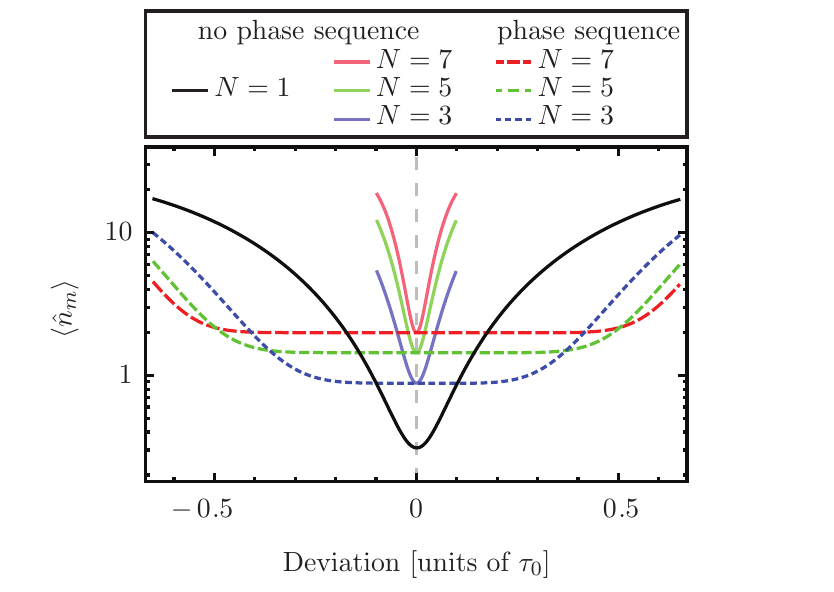}
	\caption{Phonon number expectation value after $N=1, 3, 5, 7$ interactions as a function of the deviation from the ideal interaction area. Solid curves show the phonon number when the driving phase is kept constant throughout the evolution. Dashed curves show the phonon number with a nontrivial phase sequence that minimizes the effect of deviations. In this low-loss scenario we use the driving phases from the lossless case. Physical parameters have the same values as in Fig.~\ref{Fig:MechPhase}.}\label{Fig:DeviationLongSequences}
\end{figure}

Our technique can be extended from three interactions to longer sequences with more interactions. 
The evolution under these sequences can be calculated by composing the evolution vector in Eq.~(\ref{Eq:LangevinSolution}) under successive driving phases. 
Like in the three-interaction case, phonon and photon number expectation values can be written as the sum of two contributions, a lossy oscillatory part and a thermal noise part. 
Following coherent control techniques in lossless qubits,\cite{Torosov2011a} 
we restrict ourselves to sequences with an odd number of interactions, $N$, 
where the initial and final phases are zero, 
and where the phases are symmetric around the central interaction in the composite sequence, 
\begin{eqnarray}
\phi(t) =
\left\{
\begin{array}{ccl}
0						& \hspace{0.5em}		& t\leq\tau_0				\\
\phi_2					& 						& \tau_0 < t \leq 2\tau_0	\\
\vdots					&						&							\\
\phi_{\frac{N-1}{2}}	& 						& \left( \frac{N-3}{2} \right) \tau_0 < t \leq \left( \frac{N-1}{2} \right) \tau_0	\\
\phi_{\frac{N+1}{2}}	& 						& \left( \frac{N-1}{2} \right) \tau_0 < t \leq \left( \frac{N+1}{2} \right) \tau_0	\\
\phi_{\frac{N-1}{2}}	& 						& \left( \frac{N+1}{2} \right) \tau_0 < t \leq \left( \frac{N+3}{2} \right) \tau_0	\\
\vdots																		\\
\phi_2					& 						& \left( N-2 \right) \tau_0 < t \leq \left( N-1 \right) \tau_0	\\
0						& 						& \left( N-1 \right) \tau_0 < t
\end{array}
\right.. \nonumber
\end{eqnarray}
Under these restrictions a sequence has $\frac{N-1}{2}$ free phase parameters, $\phi_2$, $\ldots$, $\phi_{\frac{N+1}{2}}$. 
These can be used to nullify the first $\frac{N-1}{2}$ derivatives of the phonon number with respect to deviations in the ideal interaction area, $\frac{\pi}{2}$. There, the phonon number is evaluated at the endtime of the sequence, $t=N\tau_0$. 
This produces a more robust optomechanical state transfer by increasing the region of parameters where the phonon number is stable against deviations, see Fig.~\ref{Fig:DeviationLongSequences}. 
The values of the phases that optimally enhance the robustness of optomechanical state transfer can be calculated numerically. 
A tradeoff of long composite sequences is that decoherence plays a more important role. 
Therefore, in high-loss scenarios, long driving sequences should be avoided.  

\subsection*{Robust transfer with smooth phase-sequences}\label{Sec:Results:Sub:Stability}
In the previous section, we discussed the effect of a constant-amplitude driving field with a time-dependent phase that is constant in the intervals $\tau_0 < t < 2\tau_0$, $2\tau_0 < t < 3\tau_0$, $\ldots$ but is discontinuous at $t=\tau_0,2\tau_0,3\tau_0,\ldots$.
This abrupt change in the driving phase can take the optomechanical system out of its semiclassical steady-state amplitude values, $\alpha$ and $\beta$. 
On the other hand, if a sufficiently-smooth phase sequence is applied, 
$\alpha$ and $\beta$ change smoothly over time and return to their steady-state values once the phase returns to its initial value. 
Suppose a phase sequence is applied between time $k \tau_0$ and $\left(k+1\right) \tau_0$. 
If the phase is sufficiently smooth and symmetric around the half sum of the initial and final times, $\left(k+1/2\right) \tau_0$, 
then the changes in the detuning, $\Delta$, and the interaction area between $k \tau_0$ and $(k+1) \tau_0$ are negligible. 

\begin{figure}[tpb!]
	\centering
	\includegraphics{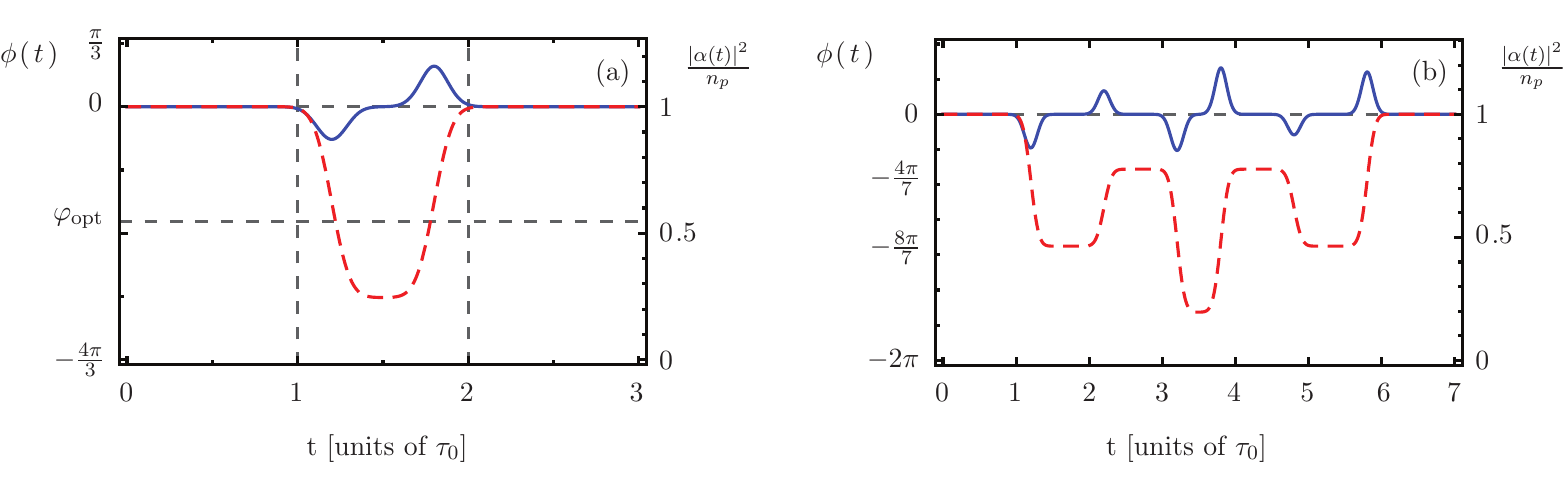}
	\caption{(a) Three- and (b) seven-interaction smooth phase sequences, $\phi(t)$ (red dashed curve), and normalized semiclassical field intensity, $\left|\alpha(t)\right|^2/n_{p}$ (blue solid curve). For both (a) and (b), the scale on the left vertical axis corresponds to the phase sequence and the scale on the right vertical axis corresponds to the normalized intensity.}\label{Fig:AlphaBetaEvolution}
\end{figure}

We verify this assertion by numerically calculating the evolution of $\alpha$ and $\beta$ driven by a smooth phase sequence for the parameters given in Fig.~\ref{Fig:Oscillations}. 
From $\tau_0$ to $2 \tau_0$ we set the phase to be given by
\begin{eqnarray}
\phi\left(t\right) = \varphi_{\mathrm{opt}} f \, \theta_s\left[t-\left(1.2\right)\tau_0,\sigma\right] \, \theta_s\left[\left(1.8\right)\tau_0-t,\sigma\right],
\end{eqnarray} 
where we assume that the driving amplitude has a smooth step function based on the error function, $\theta_s\left(t,\sigma\right) = \left[\mathrm{erf}\left(t/\sigma\right)+1\right]/2$. 
We introduce two parameters, $\sigma = 25 \, \tau_{0}$ and $f$. The former controls the smoothness of the pulse, and the latter is a dimensionless parameter of order one that keeps the average value of the phase $\phi(t)$ at its optimal value $\varphi_{\mathrm{opt}}$. In our case, the average phase between $\tau_0$ and $2 \tau_0$ is approximately $-1.89 \,\mathrm{rad}$; we used the negative value of the optimal phase in our simulations.
Following this procedure, the electromagnetic field amplitude, $|\alpha|$, smoothly decreases and increases within $8\%$ of its steady-state value. 
This leads to a change in the interaction area of less than $0.2\%$ and keeps the detuning, $\Delta$, within $0.005\%$ of its value. 
The smooth three-interaction phase sequence described in this section, and the corresponding evolution of the semiclassical field-amplitude value, $\alpha(t)$, are displayed in Fig.~\ref{Fig:AlphaBetaEvolution}(a).
We extend this calculation to a seven-interaction smooth sequence, Fig.~\ref{Fig:AlphaBetaEvolution}(b). 
The average phase values for the seven interactions are $\left\{0,-\frac{6\pi}{7},-\frac{4\pi}{7},-\frac{8\pi}{7},-\frac{4\pi}{7},-\frac{6\pi}{7},0\right\}$. 
Under this sequence, the semiclassical amplitudes remain within $10\%$ of their steady-state values, 
and these amplitudes return to their steady-state values after the sequence is over. 
Additionally, the phase sequence keeps the interaction area within $0.1\%$ of its ideal value. 
All of this can be observed in Fig.~\ref{Fig:AlphaBetaEvolution}(b).

In general, if the changes in detuning and interaction area are negligible, 
our results for discontinuous phase sequences should remain unaltered for smooth phase sequences as long as the following two conditions are fulfilled: First, the interaction area in each part of the composite sequence remains the same, $A=\pi/2$, and, 
second, the phase changes are smooth enough to keep the system in its steady state. 

\section*{Conclusions}\label{Sec:Conclusions}
Cavity optomechanical systems in the red-detuned regime display lossy oscillatory transfer between photon and phonon quantum excitations. 
We demonstrate robust optomechanical state transfer using constant-amplitude laser driving with a phase-tailored composite sequence. 

We focus on a three-interaction composite sequence, where each component produces an accumulated interaction area $\pi/2$. 
The sequence has a free phase parameter that we use to minimize the effect of deviations from the ideal behavior.
The optimal phase depends on the loss rates of the system and generalizes previous results for lossless systems. 
The dominant contribution to the phonon number is an even function of the phase. 
Consequently, two phases minimize this contribution with respect to deviations. 

We demonstrate our optimization scheme using using discontinuous piecewise phase sequences. 
Such abrupt composite sequences might take the system out of its semiclassical steady state. 
For this reason, we show that the theoretical optimal phase values can be used to generate a sufficiently-smooth phase sequence producing equivalent robust transfer. These smooth continuous sequences keep the average phase at the optimal value within each component of the sequence. They also produce negligible changes in the accumulated interaction area. 

Our method can be extended to more complicated and longer composite pulse sequences containing more free parameters. 
For an $N$-parameter sequence, the free parameters can be used to nullify the first $N$ derivatives of the final phonon number. 
This produces maximally robust transfer (in the dominant contribution) of phonon to photon excitations. 
Unlike lossless systems, the duration of phase sequences is limited by the loss rates, 
which exponentially reduce the amplitude of the oscillatory transfer. 
	
%%%%%%%%%%%%%%%%%%%%%%%%%%%%%%%%%%%%%%%%%%%%%%%%%%%%%%%%%%%%%%%%%%%%%%%%%%%%%%%%%%%%%%%%
\section*{Acknowledgments}
%%%%%%%%%%%%%%%%%%%%%%%%%%%%%%%%%%%%%%%%%%%%%%%%%%%%%%%%%%%%%%%%%%%%%%%%%%%%%%%%%%%%%%%%
C.V.V. acknowledges financial support from CONACYT through scolarship \#294810. 
B.J.A. acknowledges financial support from CONACYT through C\'{a}tedra Grupal \#551. 
E.K. acknowledges financial support from the European Union's Horizon 2020 research and innovation programme under the Marie Sk\l{}odowska-Curie grant agreement No 705256 - COPQE. 
B.M.R.-L. acknowledges financial support from 
CONACYT through CB-2015-01 \#255230 project grant; 
from Consorcio en \'Optica Aplicada through CONACYT FORDECYT \#296355 project grant 
and thanks the Photonics and Mathematical Optics Group at Tecnologico de Monterrey.

%%%%%%%%%%%%%%%%%%%%%%%%%%%%%%%%%%%%%%%%%%%%%%%%%%%%%%%%%%%%%%%%%%%%%%%%%%%%%%%%%%%%%%%%
%%%%%%%%%%%%%%%%%%%%%%%%%%%%%%%%%%%%%%%%%%%%%%%%%%%%%%%%%%%%%%%%%%%%%%%%%%%%%%%%%%%%%%%%
\section*{References}
%%%%%%%%%%%%%%%%%%%%%%%%%%%%%%%%%%%%%%%%%%%%%%%%%%%%%%%%%%%%%%%%%%%%%%%%%%%%%%%%%%%%%%%%
%%%%%%%%%%%%%%%%%%%%%%%%%%%%%%%%%%%%%%%%%%%%%%%%%%%%%%%%%%%%%%%%%%%%%%%%%%%%%%%%%%%%%%%%
%	\bibliographystyle{habbrv}
%	\bibliography{ref-robust-arXiv}

%%%%%%%%%%%%%%%%%%%%%%%%%%%%%%%%%%%%%%%%%%%%%%%%%%%%%%%%%%%%%%%%%%%%%%%%%%%%%%%%%%%%%%%%
%%%%%%%%%%%%%%%%%%%%%%%%%%%%%%%%%%%%%%%%%%%%%%%%%%%%%%%%%%%%%%%%%%%%%%%%%%%%%%%%%%%%%%%%
	
\end{document}